\documentclass[aps,pra,numerical,twocolumn,longbibliography]{revtex4-1}
\usepackage{graphicx}
\usepackage{bm}
\usepackage{amssymb}
\usepackage{multirow}
\usepackage{braket}             		
\usepackage{float}  
\usepackage{natbib}
\usepackage{cancel}

\usepackage{color}
 \usepackage{soul}

 \definecolor{darkred}{rgb}{0.8,0.1,0.1}
 \definecolor{DARKRED}{rgb}{0.8,0.1,0.1}
 \definecolor{darkblue}{rgb}{0.1,0.1,0.7}
 
\definecolor{orange}{RGB}{255,127,0}

\begin{document}
%


%
\title{Heisenberg uncertainty relations for the non-Hermitian\\resonance state solutions to the Schr\"odinger equation}
\author{Gast\'{o}n Garc\'{\i }a-Calder\'{o}n}
\email{gaston@fisica.unam.mx} \affiliation{Instituto de F\'{\i}sica,
Universidad Nacional Aut\'onoma de M\'exico, 01000 Ciudad de M\'exico, Mexico}
\author{Jorge Villavicencio}
\email{villavics@uabc.edu.mx}
\affiliation{Facultad de Ciencias, Universidad Aut\'onoma de Baja California, 22800 Ensenada, Baja California, Mexico}

\date{\today}

\begin{abstract}
Resonance (quasinormal) states correspond to non-Hermitian solutions to the Schr\"odinger equation obeying outgoing boundary conditions which lead to complex energy eigenvalues and momenta. Following the normalization rule for resonance states obtained from the residue at a complex pole of the outgoing Green's function to the problem, we propose a definition of expectation value for these states and use it to investigate the extent of validity of the Heisenberg uncertainty relations for potentials that vanish after a distance. We derive analytical expressions for the expectation values involving the momentum and the position for a given resonance state and find in model calculations that the Heisenberg uncertainty relations are satisfied for a broad range of potential parameters. A comparison of our approach with that based on the regularization method by Zel'dovich yields very similar results except for resonance energies very close to the energy threshold. Our work shows that the validity of the Heisenberg uncertainty relations may be extended to the non-Hermitian resonance state solutions to the Schr\"odinger equation.
\end{abstract}

\date{\today}

\maketitle

\section{Introduction}

Non-Hermitian solutions to the Schr\"odinger equation were already considered in the early days of quantum mechanics. As is well known, in 1928 Gamow derived the analytical expression of the exponential decay law in his studies on $\alpha$-decay by imposing outgoing (radiative) boundary conditions to the solutions of the time-dependent Schr\"odinger equation of the problem and hence to complex energy eigenvalues \cite{gamow28a,gamow28b}. Some years later, in 1939, Siegert considered outgoing boundary conditions in a time-independent framework, to show that the description of an isolated sharp resonance in the scattering of a particle by a potential of finite range could be derived from the resonant eigenstate associated with the corresponding complex energy \cite{siegert39}. In these non-Hermitian descriptions, the real part of the complex energy represents the energy of the particle and the inverse of the imaginary part is directly related to the time scale of the resonance process. However, the eigenfunctions associated with complex energy eigenvalues increase exponentially with distance and as a consequence the usual rules concerning normalization and eigenfunction expansions do not apply. Since for scattering problems the resonance eigenstates are evaluated at a finite distance the above issues did not prevent, however, the formulation of nuclear reactions based on these ideas \cite{peierls59,rosenfeld61} .

The question of the normalization of \textit{resonance states} was given considerably attention in the 1960s and 1970s. In fact, various approaches led to  similar normalization conventions. Zel'dovich proposed a regularization method \cite{zeldovich61} that was later adopted by Bergreen \cite{berggren68}. Romo obtain a normalization condition as the residue at a complex pole of the outgoing Green's function to the problem in terms of Jost functions \cite{romo68} which was shown to be identical with that obtained by Zel'dovich \cite{gyarmati71}. Another procedure referred to as the complex scaling or rotation method \cite{balslev71,giraud03} yields a similar normalization condition for elastic processes as that by Zel'dovich \cite{zeldovich61} and Romo \cite{romo68}. Here we consider the normalization condition discussed by Garc\'{\i }a-Calder\'{o}n and Peierls \cite{gcp76} which also involves the behavior of the Green's function near a complex pole but leads to a normalization rule which is given in terms of \textit{resonance  states} and goes into the usual rule in the case of bound states. This rule also extends the validity of first-order perturbation theory to  resonance  states, a point made also by Hokkyo following a different approach \cite{hokkyo65a,hokkyo65b} and, as shown below, it also coincides with the Zel'dovich prescription. The study of the analytical properties of the outgoing Green's function pave the way to derive eigenfunction expansions involving \textit{resonance  states} \cite{berggren68,romo68,more71,more73,gcp76,gc76,gareev78,tolstikhin98}.
Since the 1990's up to the present time, one may find in the literature an increasing number of  works dealing with distinct aspects of these states, as discussed, for example in Refs. \cite{mgcm05,mondragon06,rotter09,gc10,madrid12,hatano14,gcr16,gcch17} and references therein.
It is worth mentioning the generalization of the phenomenon of diffraction in time, first discussed by Moshinsky \cite{moshinsky52}, to potentials of finite range by Garc\'{\i }a-Calder\'{o}n and Rubio \cite{gcr97}. See also \cite{cgcm09}.
It is also worth mentioning work concerning \textit{resonance (quasinormal)} solutions to the Klein-Gordon equation in open electromagnetic cavities \cite{leung98} and in gravitational systems \cite{leung98,starinets09}.

Recently, it has been shown using the approach developed by Garc\'{\i }a-Calder\'{o}n, as reviewed in Refs. \cite{gc10,gc11}, that the time evolution of single particle decay by tunneling out of a potential in terms of \textit{resonance  states} and \textit{continuum wave functions} yields exactly the same result for interactions that vanish beyond a distance \cite{gcmv12}. (see also \cite{gcmv13}). An essential difference between these approaches is that the former provides analytical expressions for the distinct decaying regimes whereas the later consists of a `black-box' numerical calculation.
Since \textit{resonance  states} correspond to non-Hermitian solutions to the Schr\"odinger equation and therefore lie outside the standard formalism of quantum mechanics, the above results have prompt us to explore some fundamental issues concerning these states. One of these refers to the Born rule \cite{gcch17} and the one discussed here concerns the Heisenberg uncertainty relations.

This work, therefore, explores the Heisenberg uncertainty relations using \textit{resonance  states}. This requires to define the expectation value of a given operator in terms of these states. This issue was addressed by a number of authors some decades ago \cite{hokkyo65a,romo68,gyarmati72,berggren96}.  We shall not be concerned here with the complex scaling method \cite{burgers96,moiseyev98,papadimitriou16} mainly because it requires distinct mathematical considerations. In any case, as far as we know, the study of the Heisenberg uncertainty relations involving \textit{resonance} states has not been addressed before.

The paper is organized as follows: In Sec. \ref{main} we review some relevant properties of \textit{resonance states}. Section \ref{expectation} provides both our definition for the expectation value and that involving the regularization method in terms of resonance states and gives the corresponding Heisenberg uncertainty relations. In Sec. \ref{models} we illustrate our results using two solvable models and we end with the conclusions in Sec. \ref{conclusions}.

\section{Resonance states \label{main}}

We shall be concerned here with the time-honored problem of a single particle subjected to a spherically symmetric real potential that vanishes exactly beyond a distance, i.e., $V(r)=0,\, r >a$. For simplicity of the discussion we shall refer to the case of zero angular momentum, though the extension to higher angular momentum is straightforward. Our analysis will be therefore equivalent to a description on the half-line in one dimension.

\textit{Resonance (quasinormal)} states obey  the Schr\"odinger equation to the problem with the complex energy
eigenvalues \cite{gamow28a,gamow28b,siegert39,peierls59,rosenfeld61}. Using units $\hbar=2m=1$ we may write,
\begin{equation}
[E_n-H]u_n(r)=0,
\label{f1}
\end{equation}
where $E_n=k_n^2=\mathcal{E}_n-i\Gamma_n/2$. Since $k_n=\alpha_n-i\beta_n$, then $\mathcal{E}_n=\alpha_n^2-\beta_n^2$ and
$\Gamma_n=4\alpha_n\beta_n$. The Hamiltonian $H$ in (\ref{f1}) reads,
\begin{equation}
H=-\frac{d^2}{dr^2} + V(r).
\label{f1a}
\end{equation}
The solutions $u_n(r)$ to Eq. (\ref{f1}) vanish at the origin and satisfy outgoing (radiative) boundary conditions, namely,
\begin{equation}
u_n(0)=0, \qquad \left [ \frac {d}{dr}u_n(r) \right]_{r=a_-}= ik_n\,u_n(a).
\label{f2}
\end{equation}
The expression on the right-hand side of (\ref{f2}) follows from the fact that for $r >a$,
\begin{equation}
u_n(r)=D_ne^{ik_nr}=D_ne^{i\alpha_n r}ê^{\beta_n r},
\label{f2a}
\end{equation}
which yields a divergent result for the usual normalization rule.
Due to time reversal invariance, Eq. (\ref{f1}) admits also the solutions $k_{-n}=-k_n^*$ and $u_{-n}(r)=u_n^*(r)$ \cite{rosenfeld61}. It turns out that the set of complex values $\{k_n\}$ corresponds to the poles the outgoing Green's function to the problem, which are the same as those of the corresponding $S(k)$ matrix. They are distributed in the complex $k$ plane in a well known manner and are simple except in special cases. We assume that such is the case here. In general they lie either on the positive imaginary axis, corresponding to bound states, or in the lower half of the $k$ plane, corresponding to antibound states (lying on the negative imaginary $k$ axis) and to the infinite set of resonance  states
\cite{newtonchap12}.

As mentioned above, another form to define \textit{resonance  states}, is from the residue $\rho_n(r,r')$ at a given pole $k_n$  of the outgoing Green's function to the problem derived in Ref. \cite{gcp76} in the energy plane adapted for the $k$ plane \cite{gc10}, namely,
\begin{equation}
\rho_n(r,r')=\frac{u_n(r)u_n(r')}{2k_n \{\int_0^a\,u_n^2(r)\,dr+ (i/2k_n)u_n^2(a)\}}.
\label{f3}
\end{equation}
It is worthwhile to emphasize that the set of poles $\{k_n\}$, and hence the states $\{u_n(r)\}$, vary as a function of the parameters that define the potential. This means that each pole follows a trajectory along the complex $k$ plane as a function of the potential parameters and hence, as noticed by Nussensveig many years ago, a complex pole may become a bound state or viceversa \cite{nussenzveig59}. For bound states, i.e., $k_b=i\gamma_b$, with $\gamma_b$ a real positive  number, the expression within brackets in (\ref{f3}) yields exactly the the usual  normalization rule of a bound state,
namely,
\begin{equation}
\frac {i}{2k_b}u_b^2(a) = \frac {1}{2\gamma_b}u_b^2(a)\equiv \int_a^\infty u_b^2(r)\,dr,
\label{f3a}
\end{equation}
where $u_b(r)=D_b\exp(-\gamma_b r)$. This suggests to adopt in general the normalization rule for \textit{resonance  states} \cite{gcp76,gc10} as,
\begin{equation}
\int_0^au_n^2(r)dr + \frac{i}{2k_n}u_n^2(a)=1.
\label{f4}
\end{equation}
It might be of interest to add that the normalization condition given by (\ref{f4}) is physically satisfactory in that it extends the validity of first-order perturbation theory  for a change in the internal potential  of the system, by which the change in the energy eigenvalue is given as the integral of the perturbing potential times the square of the wave function \cite{gcp76}.

It might be also of interest to refer to some sort of orthonormality condition concerning resonance states.
Using Green's formula between Eq. (\ref{f1}) times $u_m(r)$ and similar equation for $u_m(r)$ times $u_n(r)$, subtracting and integrating the resulting expression along the internal interaction region and using the corresponding boundary conditions for $u_n$ and $u_m$, allows us to write, in view of (\ref{f4}), the orthonormality condition \cite{gc10},
\begin{equation}
\int_0^au_n(r)u_m(r)\,dr + \frac{i}{k_n+k_m}u_n(a)u_m(a)=\delta_{nm}.
\label{f5}
\end{equation}

Another useful expression which follows from the expansion of the outgoing Green's function in terms of \textit{resonance} states \cite{gc10},
\begin{equation}
G^+(r,r';k)=\sum_{n=-\infty}^{\infty} \frac{u_n(r)u_n(r')}{2k_n(k-k_n)}, \qquad (r,r')^\dagger \leq a,
\label{f4aa}
\end{equation}
is the closure relation \cite{gc10},
\begin{equation}
\frac{1}{2} \sum_{n=-\infty}^{\infty}\, u_n(r)u_n(r') = \delta(r-r'), \qquad (r,r')^\dagger \leq a,
\label{f6}
\end{equation}
where the notation $(r,r')^{\dagger} \leq a$ means that the above expressions hold along the internal interaction region and at the boundary values except when $r=r'=a$.

Using (\ref{f6}) one may immediately expand an arbitrary function $\Psi(r)$ defined in the interval $(0,a)$ as,
\begin{equation}
\Psi(r)= \frac{1}{2} \sum_{n=-\infty}^{\infty}\, C_nu_n(r), \quad r\leq a,
\label{f8}
\end{equation}
where the coefficient $C_n$ reads,
\begin{equation}
C_n= \int_0^a \Psi(r)u_n(r)\,dr.
\label{f9}
\end{equation}
The time dependence of a resonance state is given by
\begin{equation}
u_n(r,t)=u_n(r)e^{-i\mathcal{E}_n t}e^{-\Gamma_n t/2},
\label{2cc}
\end{equation}
which shows that the amplitude of a resonant state decreases exponentially with time.
Since the absolute value square of the resonance function varies with time, it is of interest to see the result of applying
the continuity equation, or more precisely of the continuity equation integrated along the internal region of the interaction,
\begin{equation}
\frac{\partial}{\partial t} \int_0^a |u_n(r,t)|^2 \,dr =-2 \textrm{Im} \left [u_n^*(r,t)\frac{\partial }
{\partial r} u_n(r,t)\right ]_0^a.
\label{2e}
\end{equation}
Now, using in the above expression (\ref{2cc}) and the boundary conditions given by (\ref{f2}) allows us to write,
\begin{equation}
\Gamma_n= 2\alpha_n \frac{|u_n(a)|^2}{\int_0^a|u_n(r)|^2\,dr},
\label{2f}
\end{equation}
which establishes that the decay width is proportional to the velocity of the decaying particle times the probability to find the particle at the surface divided by the probability to find it inside the interaction region. It is worth noticing that (\ref{2f}) depends on the absolute value squared $|u_n(r)|^2$ whereas  the normalization condition given by (\ref{f4}) depends on $u_n^2(r)$.

Since one of the model calculations is in one dimension, we end this Section by mentioning that in one dimension the outgoing boundary condition occur at both end points of the system and hence, the normalization condition involves an additional surface term. However, the expressions for the resonance expansions and the closure relationship remain the same \cite{gcr97,gc10}.

\section{Expectation values \label{expectation}}

In analogy with the normalization condition (\ref{f4}), one could define the expectation value of an operator $\mathcal{O}$ in terms of \textit{resonance states} as,
\begin{equation}
\braket{\mathcal{O}}= \int_0^a\, u_n(r)\mathcal{O}u_n(r)\,dr + \frac{i}{2k_n}\left [u_n(r) \mathcal{O} u_n(r) \right]_{r=a}.
\label{e1}
\end{equation}
A similar definition has been also given by Hokkyo \cite{hokkyo65a,hokkyo65b} and also in the definition of internal product involving the Klein-Gordon equation by Leung et.al.  \cite{leung98}. Clearly, the definition given by (\ref{e1}) is unsuitable on physical grounds since it yields a complex quantity. For example, choosing in (\ref{e1}) $\mathcal{O}=H$, the Hamiltonian to the system,  using (\ref{f4}), yields by substitution in (\ref{e1}),
\begin{equation}
\braket{H}=\mathcal{E}_n -i \frac{1}{2}\Gamma_n.
\label{e3}
\end{equation}
Equation (\ref{e3}) suggests, however, that the real part of (\ref{e3}) yields the correct answer and hence it
prompts to define the expectation value of an operator in a given \textit{resonance  state} as the real part of the (\ref{e1}), namely,
\begin{equation}
\braket{\braket {\mathcal{O}}} \equiv {\rm Re}\,\braket{\mathcal{O}}.
\label{e2}
\end{equation}
Clearly, for $\mathcal{O}=H$, we obtain
\begin{equation}
\braket{\braket {\mathcal{H}}} =\mathcal{E}_n.
\label{e3n}
\end{equation}

Since our  aim is to calculate the Heisenberg uncertainty relations, it is required to calculate Eq. (\ref{e2}) for $\mathcal{O}=p, p^2,r$ and $r^2$.

For $\mathcal{O}=p=-id/dr$, integrating by parts the integral term in (\ref{e1}), and using the boundary condition at $r=a$ given in (\ref{f2}) to calculate the surface term in (\ref{e1}), one obtains
\begin{equation}
\braket{\braket{p}}= 0.
\label{e7}
\end{equation}
For $\mathcal{O} = p^2=-d^2/dr^2$, one may use the identity $p^2=H-V(r)$, that follows from (\ref{f1a}), to write using (\ref{e3}),
\begin{equation}
\braket{\braket{p^2}}= \mathcal{E}_n - {\rm Re}\,\left \{\int_0^a\,V(r)u_n^2(r)\,dr \right \}.
\label{e8}
\end{equation}

It is worth to discuss a relevant implication of Eq. (\ref{e8}). One should notice, since $k_n=\alpha_n-i\beta_n$ and hence $\mathcal{E}_n=k_n^2=\alpha_n^2 -\beta_n^2$, that in the limit of a vanishing potential the imaginary part of the complex poles goes to $\infty$, namely,
\begin{equation}
k_n \to \alpha_n - i \infty,
\label{e8a}
\end{equation}
as expected for the free case where complex poles are absent.
It follows then by inspection of (\ref{e8}), that in general there might be potentials, usually very shallow potentials, where for a given $u_n(r)$, $\beta_n > \alpha_n$, and hence, $\braket{\braket{p^2}} < 0$, which seems unacceptable on physical grounds. Hence, in order to have $\braket{\braket{p^2}} > 0$, the potential parameters must guarantee that the complex poles fulfill  $\alpha_n > \beta_n$. These poles are called \textit{proper poles} and are usually the case for most problems of physical interest.

Using Eqs. (\ref{e1}) and (\ref{e2})  for $\mathcal{O}= r$ and respectively for  $r^2$ leads to the expressions,
\begin{equation}
\braket{\braket{r}}= {\rm Re} \left \{ \int_0^a\, ru_n^2(r)\,dr + \frac{i}{2k_n}au_n^2(a) \right \},
\label{e5}
\end{equation}
and
\begin{equation}
\braket{\braket{r^2}}= {\rm Re} \left \{\int_0^a\, r^2u_n^2(r)\,dr + \frac{i}{2k_n}a^2u_n^2(a) \right \}.
\label{e6}
\end{equation}
For bound states, however,  since $k_b=i\gamma_b$, the integral over the full space is well defined and hence one may write the expectation value in the usual way.

\subsection{The regularization procedure\label{zeldovich}}

In 1961 Zel'dovich suggested a regularization procedure to obtain a normalization rule for \textit{resonance  states}. This procedure consists in introducing a factor $\exp(-\varepsilon r^2)$ into the corresponding integrand of the usual normalization condition and subsequently taking the limit as $\varepsilon \to 0$, namely,
\begin{equation}
\lim_{\varepsilon\to 0}\int_0^{\infty} e^{-\varepsilon r^2}\,u_n^2(r)\,dr =1.
\label{b1}
\end{equation}
In 1968, Berggren adopted the  normalization procedure of Zel'dovich  and generalize it to define the expectation value of an operator $\mathcal{O}$ as,
\begin{equation}
\braket{\mathcal{O}}_{B}=\lim_{\varepsilon\to 0}\int_0^{\infty} e^{-\varepsilon r^2}\,u_n(r)\mathcal{O}\,u_n(r)\,dr.
\label{b2}
\end{equation}
Noticing, however,  that (\ref{b2}) yields a complex quantity, Berggren suggested to define the expectation value of an operator as \cite{berggren68},

\begin{equation}
\braket{\braket{\mathcal{O}}}_{B}={\rm Re}\braket{\mathcal{O}}_{B},
\label{b4}
\end{equation}
which yields the same result for ${\mathcal{O}}=H$  as that given by  Eq. (\ref{e3n}), namely \cite{berggren68},
\begin{equation}
\braket{\braket {\mathcal{H}}}_{B} =\mathcal{E}_n.
\label{b4a}
\end{equation}

Since the potential vanishes exactly beyond the distance $a$,  it is convenient to write (\ref{b2}) as,
\begin{eqnarray}
\braket{\mathcal{O}}_{B}&=&\int_0^{a} u_n(r)\mathcal{O}u_n(r)\,dr +\nonumber \\ [.3cm]
&&\lim_{\varepsilon\to 0}\int_a^{\infty} e^{-\varepsilon r^2}u_n(r)\mathcal{O}u_n(r)\,dr.
\label{b3}
\end{eqnarray}

As shown in appendix \ref{appendixA}, the regularization procedure yields,
\begin{equation}
\braket{\braket{p}}_{B}=0,
\label{b11}
\end{equation}
and
\begin{equation}
\braket{\braket{p^2}}_B= \mathcal{E}_n - {\rm Re}\,\left \{\int_0^a\,V(r)u_n^2(r)\,dr \right \},
\label{b12}
\end{equation}
which are results identical to those given by (\ref{e7}) and (\ref{e8}).

Let us now refer to the case $\mathcal{O}=r^m$. Using (\ref{f2a}) we may write (\ref{b2}) as
\begin{equation}
\braket{r^m}=\int_0^{a} r^m\,u_n^2(r)\,dr
+\lim_{\varepsilon\to 0}D_n^2\int_a^{\infty} e^{-\varepsilon r^2}\,r^m\,e^{z r}\,dr,
\label{b5}
\end{equation}
where we have defined $z=2ik_n$ and $D_n$ stands for the normalization coefficient of the resonance state. The second integral in Eq. (\ref{b5}) may be calculated with the help of the identity given  Eq. (\ref{identity2}) of  Appendix \ref{appendixA},  to get,
\begin{equation}
\braket{r^m}=\int_0^{a} r^m\,u_n^2(r)\,dr-
D_n^2\frac{\partial^m}{\partial z^m} \left[\frac{e^{za}}{z} \right].
\label{b6}
\end{equation}
Then by substituting the variable $z=2ik_n$ in (\ref{b6}) we obtain the expression,
\begin{equation}
\braket{r^m}=\int_0^{a} r^m\,u_n^2(r)\,dr+D_n^2\frac{1}{(2i)^m}\frac{\partial^m}{\partial k_n^m} \left[\frac{i}{2k_n}e^{2 i k_n a}\right].
\label{b7}
\end{equation}

Recalling that $u_n^2(a)=D_n^2 e^{2ik_na}$, we may verify using (\ref{b7}), that the case  with $m=0$ corresponds to the normalization condition given by (\ref{f4}), namely,
\begin{equation}
\lim_{\varepsilon\to 0} \int_0^{\infty} e^{-\varepsilon r^2}\,u_n^2(r)\,dr
=\int_0^{a}\,u_n^2(r)\,dr+\frac{i}{2k_n}u_n^2(a)=1,
\label{b8}
\end{equation}
which shows that the normalization procedure given by (\ref{f4}) and that by Zel'dovich  are equivalent.
Using Eq. (\ref{b7}) with with $m=1$ yields, using (\ref{b4}),
\begin{equation}
\braket{\braket{r}}_{B}={\rm Re}\left \{\int_0^{a} r\,u_n^2(r)\,dr+\frac{i}{2k_n}au_n^2(a)\left [1+\mathcal{A}_n(a)\right ] \right \},
\label{b9}
\end{equation}
where
\begin{equation}
\mathcal{A}_n(a)=\frac{i}{2k_na}.
\label{b9a}
\end{equation}
In a similar fashion, using Eq. (\ref{b7}) with $m=2$, yields,
\begin{equation}
\braket{\braket{r^2}}_{B}= {\rm Re}\left \{\int_0^{a} r^2\,u_n^2(r)\,dr+
\frac{ia^2}{2k_n}u_n^2(a)\left[1 +\mathcal{B}_n(a) \right]\right \},
\label{b10}
\end{equation}
where,
\begin{equation}
\mathcal{B}_n(a)=\frac{i}{k_na}-\frac{1}{2(k_na)^2}.
\label{b10a}
\end{equation}

It turns out, that for a bound state, $k_n=i\gamma_n$, the expectation values for $r$ and $r^2$, given respectively by (\ref{b9}) and (\ref{b10}), give the correct answer. This may be verified by direct partial integration since in this case it is not necessary  to make use of  the regularization procedure. However, it follows by inspection of the corresponding terms, given by (\ref{b9a}) and (\ref{b10a}), that the resulting contributions are in general very small unless $\gamma_n a \ll 1$. Since this may happen in general for a bound state very close to the energy threshold, one may expect that in general these contributions will be negligible. A similar argument may be employed for a resonance state very close to the energy threshold, namely,  $|k_n a| \ll 1$.

Summarizing, one sees  that  (\ref{e2}) and (\ref{b4}) yield identical results, in addition to the normalization condition,  for $\mathcal{O}=H, p, p^2$. On the other hand, we find that for  ${\mathcal{O}}=r, r^2$  both procedures yield different results. However, since in general $|k_n a| >> 1$, one may expect the  terms $\mathcal{A}_n(a)$ and $\mathcal{B}_n(a)$ to provide small corrections to $\braket{\braket{r}}$ and $\braket{\braket{r^2}}$.

\subsection{Heisenberg uncertainty relations}

Using the expression for the expectation value of an operator $\mathcal{O}$ given by  (\ref{e2}), the corresponding dispersion, defined in the usual way, reads
\begin{equation}
\Delta \mathcal{O}= \sqrt{[\braket{\braket{\mathcal{O}^2}}-\braket{\braket{\mathcal{O}}}^2]}.
\label{ia1}
\end{equation}

It follows from (\ref{ia1}) that the expression for the Heisenberg uncertainty relations $(\Delta r) (\Delta p)$, in view of (\ref{e7}) may be written as,
\begin{equation}
\Delta r \Delta p=\sqrt{\left [\braket{\braket{r^2}}-\braket{\braket{r}}^2\right ]
\braket{\braket{p^2}}},
\label{i1}
\end{equation}
which may be evaluated using Eqs. (\ref{e8}), (\ref{e5}), and (\ref{e6}).

In a similar fashion, for the regularization procedure,
\begin{equation}
[\Delta r \Delta p]_{B}=\sqrt{\left [ \braket{\braket{r^2}}_{B}-\braket{\braket{r}}^{2}_{B}\right ]
\braket{\braket{p^2}}_{B}},
\label{i1B}
\end{equation}
which may be evaluated using Eqs. (\ref{b12}), (\ref{b9}), and (\ref{b10}).

As is well known, the Heisenberg uncertainty relations (recalling that in our units $\hbar=1$), must satisfy
\begin{equation}
\Delta r \Delta p \geq \frac{1}{2},
\label{i2}
\end{equation}
and similarly, for the regularization procedure,
\begin{equation}
[\Delta r \Delta p]_{B} \geq \frac{1}{2}.
\label{i3}
\end{equation}
\subsubsection*{One dimension}

As pointed out above, in one dimension the potential has an additional endpoint, i.e., $V(x)=0$ for $x<0$ and $x>L$. It may be shown that the corresponding normalization condition reads \cite{gcr97},
\begin{equation}
\int_0^Lu_n^2(x)dx + \frac{i}{2k_n}[u_n^2(0)+u_n^2(L)]=1,
\label{od1}
\end{equation}
and,
\begin{eqnarray}
&&\int_0^Lu_n(x)u_m(x)dx +   \nonumber \\ [.3cm]
&&\frac{i}{k_n+k_m}[u_n(0)u_m(0)+ u_n(L)u_m(L)]=\delta_{nm}.
\label{od2}
\end{eqnarray}
In a similar fashion as discussed for the three dimensional case, the expectation value of an operator $\mathcal{O}$ now reads,
\begin{eqnarray}
&&\braket{\mathcal{O}}= \int_0^L\, u_n(x)\mathcal{O}u_n(x)\,dx + \nonumber \\ [.3cm]
&&\frac{i}{2k_n}\left [u_n(x) \mathcal{O} u_n(x) \right]_{x=0} +\frac{i}{2k_n}\left [u_n(x) \mathcal{O} u_n(x) \right]_{x=L}.
\label{od3}
\end{eqnarray}
As a consequence, the expectation values of $x$ and $x^2$ look similar to those given by Eqs. (\ref{e5}) and (\ref{e6}) except that each of them has an additional surface term. Also, using Eq. (\ref{od2}), those of $p$ and $p^2$ look similar to Eqs. (\ref{e7}) and (\ref{e8}).

Similarly, for the regularization method in one dimension the expressions for $\braket{x^m}_B$ and $\braket{p^m}_B$ read,
\begin{eqnarray}
\braket{x^m}_B&=& \int_0^{L} x^m u_n^2(x)dx \nonumber \\ [.3cm]
&+&\frac{1}{(2 i)^m} A_n^2 \frac{\partial^m}{\partial k_n^m} \left( \frac{i}{2k_n}e^{2 i k_n L} \right)  \nonumber \\ [.3cm]
&+& (-1)^m u_n^2(0) \frac{1}{(2 i)^m} \frac{\partial^m}{\partial k_n^m} \left( \frac{i}{2k_n} \right),
\label{1Dxm}
\end{eqnarray}
where $u_n(L)=A_ne^{i k_n L}$, and,
\begin{eqnarray}
\braket{p^m}_B &=& \frac{1}{i^m}\int_0^{L} u_n(x) \frac{d^mu_n(x)}{dx^m} \,\,dx \nonumber \\ [.3cm]
&+&\frac{i}{2} \,k_n^{m-1}\,\left[u_n^2(L)+(-1)^m\,u_n^2(0) \right].
\label{1Dpm}
\end{eqnarray}
From the above equation it follows that $\braket{p}_B=0$ and
\begin{equation}
\braket{p^2}_B=E_n- \int_0^{L}V(x)\,u_n^2(x) \,\,dx,
\label{pcuadr1D}
\end{equation}
which are similar to the result in three dimensions.

\section{Models\label{models}}

In order to investigate to what extent, if any, the Heisenberg uncertainty relations are satisfied  using \textit{resonance  states}, we discuss below two potential models, the  $s$-wave $\delta$-shell potential  in three dimensions and the rectangular barrier in one dimension.

\subsection{Delta-shell potential}

The s-wave $\delta$-shell potential constitutes an exactly solvable model which allows us to calculate the corresponding  \textit{resonance states} and complex poles. This is a model that has been widely used in scattering and decay problems. A nice feature of this model is that for $\lambda \to \infty$, the region $ 0 < r < a$, becomes the well known problem of a box with an infinite wall, which corresponds to an Hermitian problem. A nice feature of this model is that it permits to analyze how  a non-Hermitian open system becomes an Hermitian closed system. This potential is defined as,
\begin{equation}
V(r)=\lambda\delta(r-a),
\label{eq}
\end{equation}
where  $\lambda$ refers to the the intensity of the potential.

The \textit{resonance  state} solutions to Eq. (\ref{f1}) with the potential given by (\ref{eq}) read,
\begin{equation}
u_n(r)=\left\{
\begin{array}{cc}
A_n \,\sin (k_nr) & r\leq a \\[.3cm]
D_n \,e^{ik_nr}, & r \geq a.
\end{array}
\right.
\label{5c}
\end{equation}
From the continuity of the above solutions and the discontinuity of its derivatives with respect to $r$ (due to the $\delta$-function interaction) at the boundary value $r=a$, it follows that the complex $k_n$'s satisfy the equation,
\begin{equation}
J(k_n)=2ik_n + \lambda ( e^{2ik_na}-1)=0.
\label{5d}
\end{equation}
For $\lambda > 1$ one may write approximate analytical solutions to Eq. (\ref{5d}) as \cite{gc10,gc11},
\begin{equation}
k_n \approx \frac{n\pi}{a} \left (1-\frac{1}{\lambda a}\right ) -i \,\frac{1}{a}\left( \frac{n\pi}{\lambda a}\right )^2.
\label{5e}
\end{equation}
As is well known, using the  above expression as an initial value in the Newton-Rapshon iteration method,
\begin{equation}
k_n^{r+1}= k_n^r - \frac{J(k_n^r)}{\dot{J}(k_n^r)},
\label{5ea}
\end{equation}
where $\dot{J}=[dJ/dk]_{k=k_n}$, one may obtain the $k_n$'s with the desired degree of approximation.

As $\lambda \to 0$, one may also obtain by manipulating (\ref{5d}) that,
\begin{equation}
k_n =  \left (n-\frac{1}{2}\right )\frac{\pi}{a} -i \infty.
\label{5ee}
\end{equation}
The above limit holds also for a potential barrier of height $V_0$ and width $a$ \cite{nussenzveig72p}, both cases in agreement with Eq. (\ref{e8a}).

The normalization coefficient for resonant states may be evaluated by substitution of Eq. (\ref{5c}), for $r \leq a$, into Eq. (\ref{f4}), to obtain the analytical expression,
\begin{equation}
A_n=\left [\frac{2\lambda }{\lambda a+ e^{-2i k_n a}} \right ]^{1/2},
\label{5f}
\end{equation}
and similarly from the continuity of the solutions $u_n(r)$  at $r=a$,
\begin{equation}
D_n=-\left [\frac{2\lambda }{\lambda a+ e^{-2i k_n a}} \right ]^{1/2} \frac{k_n}{\lambda} e^{-2ik_na}.
\label{5g}
\end{equation}
\subsubsection*{Infinite wall potential}

As mentioned above, as  $\lambda \to \infty$, the system becomes a closed system whose solutions are well known in introductory courses in Quantum Mechanics. Indeed from (\ref{5e}), (\ref{5f}) and (\ref{5g})  one sees immediately that in the above limit
\begin{equation}
k_{\pm n} \to \pm \frac{n\pi}{a},
\label{iw1}
\end{equation}
and
\begin{equation}
A_n \to (2/a)^{1/2}, \qquad D_n \to 0.
\label{iw2}
\end{equation}
As a result, the non-Hermitian resonance  solutions $u_n(r)$, given by  (\ref{5c}), go into the well known infinite wall Hermitian solutions $\phi_n(r)$ ,
\begin{equation}
\phi_{\pm n}(r)=\pm \left (\frac{2}{a} \right )^{1/2} \sin \left [\frac{n\pi}{a}r \right],
\label{i4}
\end{equation}
with $n=1,2,3,... $. Notice, since $\phi_{-n}(r)= -\phi_{n}(r)$, that in the limit $\lambda \to \infty$, the sum rule given by Eq. (\ref{f6}) becomes de usual relation,
\begin{equation}
\sum_{n=1}^{\infty}\, \phi_n(r)\phi_n(r') = \delta(r-r'), \quad 0 < (r,r') < a.
\label{i4a}
\end{equation}

As is well known from quantum mechanics textbooks, the uncertainty relations for the infinite wall model read,
\begin{equation}
[(\Delta r) \,(\Delta p)]_{iw}=\sqrt{\frac{n^2\pi^2}{12}-\frac{1}{2}}.
\label{i5}
\end{equation}
\begin{figure}[htbp!]
\includegraphics[width=3.6in]{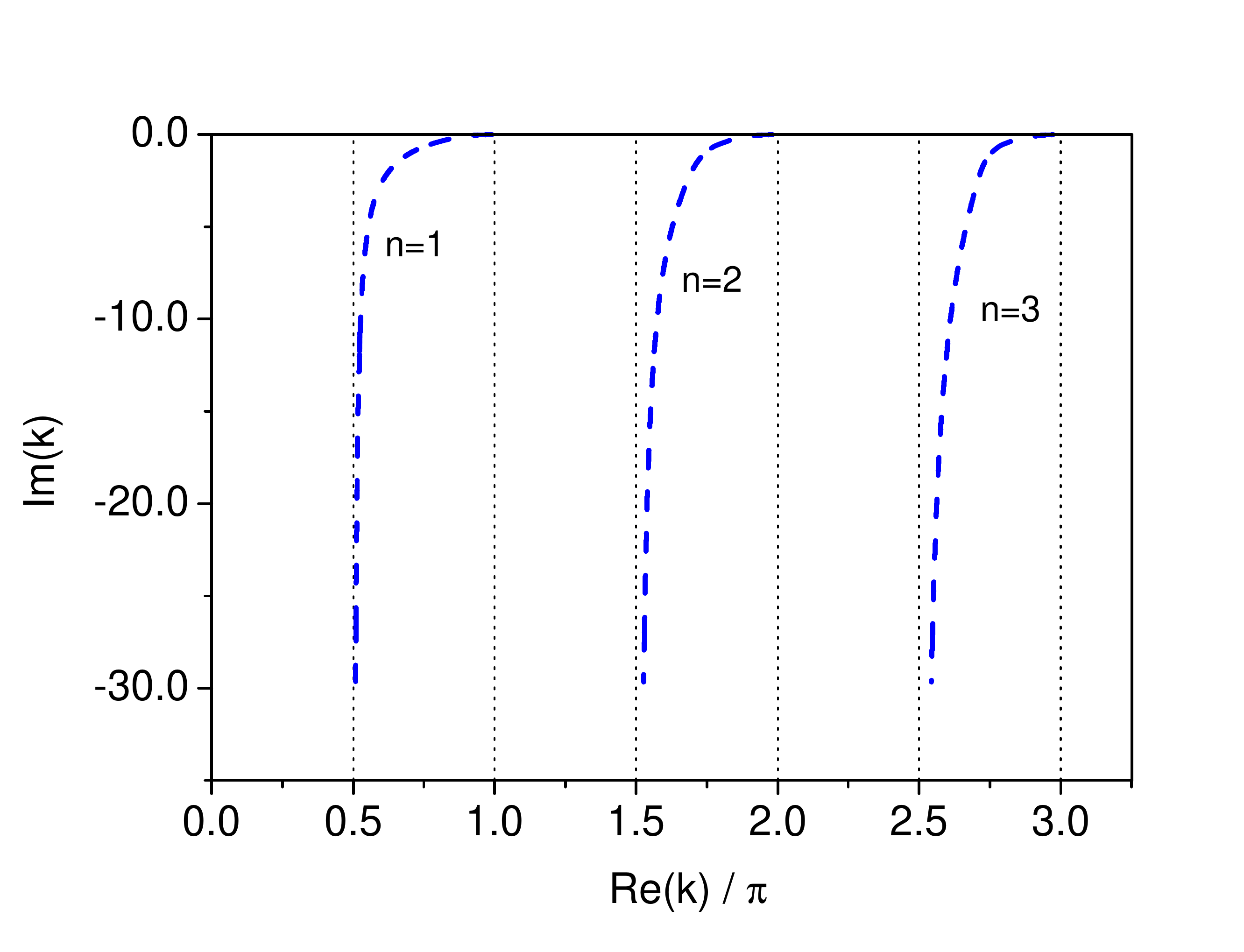}
\caption{ Plot of the first three poles $k_n=\alpha_n-i\beta_n$ on the fourth quadrant of the complex $k$ plane as a function of the intensity $\lambda$ of the $\delta$-shell potential given by Eq. (\ref{eq}) with $a=1$. For $\lambda \to \infty$ the poles $k_n$ go, as described by Eq. (\ref{5e}), into the infinite box model solutions and as $\lambda \to 0$ the corresponding imaginary parts $\beta_n$ go to $-\infty$, as described by Eq. (\ref{5ee}). See text.}
\label{Fig1}
\end{figure}

Let us know discuss the results of our calculations.

Figure \ref{Fig1} exhibits a plot of the first three complex poles $k_n$, with $n=1,2,3$, on the fourth quadrant of the complex $k$ plane. We recall that the poles $k_{-n}$ located on the third quadrant of the $k$ plane fulfill, from time reversal considerations, that $k_{-n}=-k_n^*$. As pointed out above, as $\lambda \to \infty$ the complex poles $k_n$ become real and go into the infinite wall box model solutions (\ref{iw1}), whereas, as $\lambda \to 0$, the imaginary part of the poles, $\beta_n \to -\infty$ in agreement with Eq.  (\ref{5ee}).

\begin{figure}[htbp!]
\includegraphics[width=3.6in]{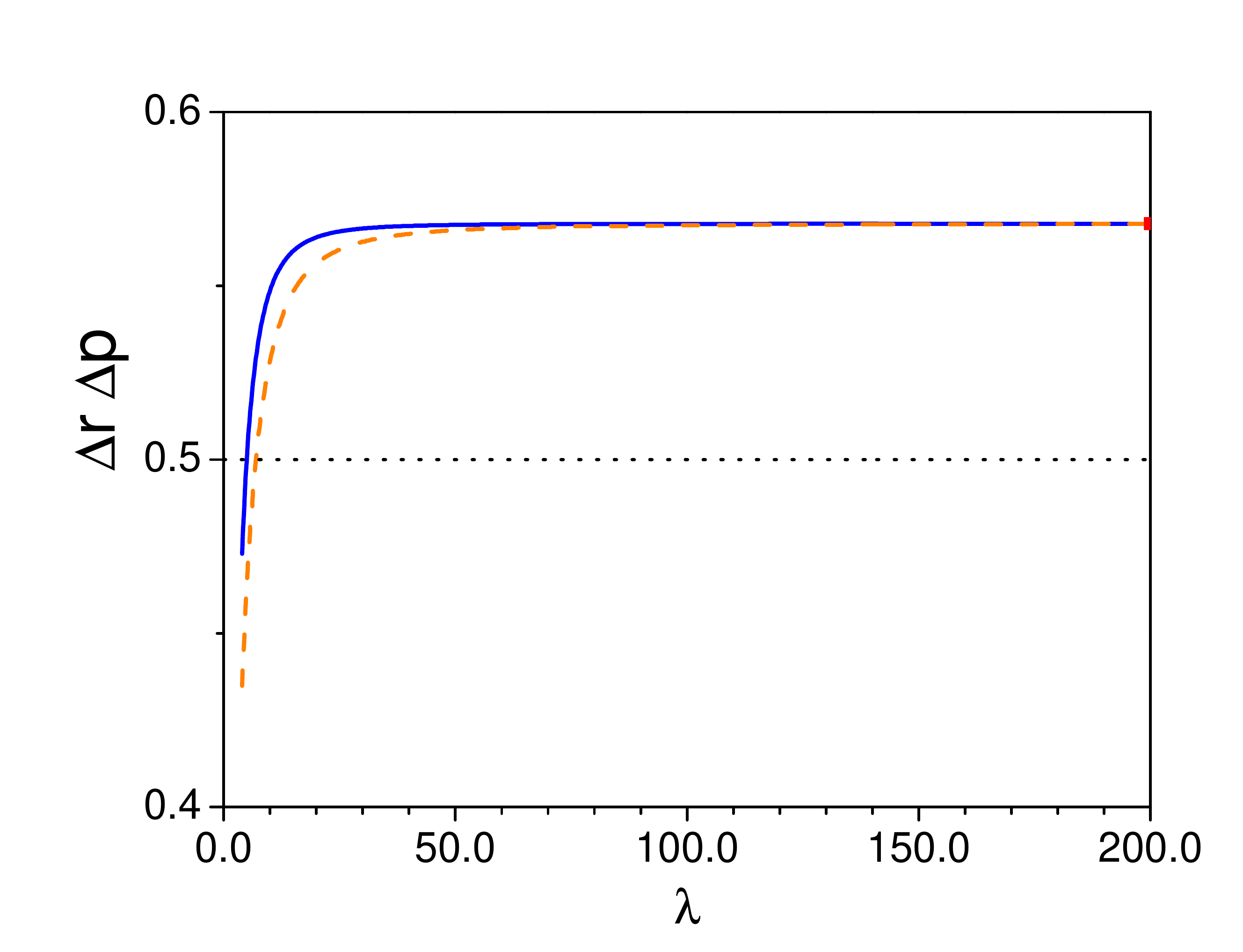}
\caption{ Plot of the uncertainty $\Delta r \Delta p$  for a $\delta$-shell  potential  as a function of the intensity $\lambda$, using Eqs. (\ref{i1}) (blue solid line), and (\ref{i1B}) (orange dashed line) for the case  $n=1$.  The uncertainty $\Delta r \Delta p$  for the infinite well ($\lambda \to \infty$) (red solid square) (Eq. (\ref{i5}) with $n=1$) is  also included for comparison. See text.}
\label{Fig2}
\end{figure}

In Fig. \ref{Fig2} we plot the Heisenberg uncertainty relations $\Delta r \,\Delta p$ as function of $\lambda$ to make a comparison between our prescription, given by (\ref{i1}) (blue solid line), and the regularization procedure, given by (\ref{i1B}) (orange dashed line). We consider the case $n=1$, and as may be appreciated, the Heisenberg uncertainty relationships obey $\Delta r \,\Delta p \ge \frac{1}{2}$  for a broad range of values of $\lambda$. For $\lambda \gtrsim 7$ both (\ref{i1}) and (\ref{i1B}) are indistinguishable, which implies that the corrections  $\mathcal{A}_n(a)$ and $\mathcal{B}_n(a)$, given respectively by (\ref{b9a}) and (\ref{b10a}), are negligible. The corresponding value of $\Delta r \,\Delta p$ for the infinite wall, computed using Eq. (\ref{i5}) for $n=1$, is indicated with a red solid square in Fig. \ref{Fig2}. It is worth noticing that as $\lambda$ increases, $\Delta r \,\Delta p$ approaches the value corresponding to the infinite wall model. Notice also,  that the uncertainty relations are not fulfilled for small values of $\lambda$. For example, $\lambda \lesssim 5$, for the case of Eq. (\ref{i2}), and  $\lambda \lesssim 7$, for the case of Eq. (\ref{i3}).

\subsection{Rectangular potential}

Here we consider the well known rectangular potential in one dimension, defined as,
\begin{equation}
V(x)=\left\{
\begin{array}{cc}
V_0, & \,\,\, 0 \leq x \leq L \\[.3cm]
0, \, &\,\, x < 0,\quad  x > L,
\end{array}
\right.
\label{1rp}
\end{equation}
where $V_0$ is a constant representing  the barrier height and $L$ is the corresponding barrier width.
As is the case for any finite range potential, the rectangular potential possesses an infinite number of complex poles $\{k_n\}$.
\begin{figure}[htbp!]
\includegraphics[width=3.6in]{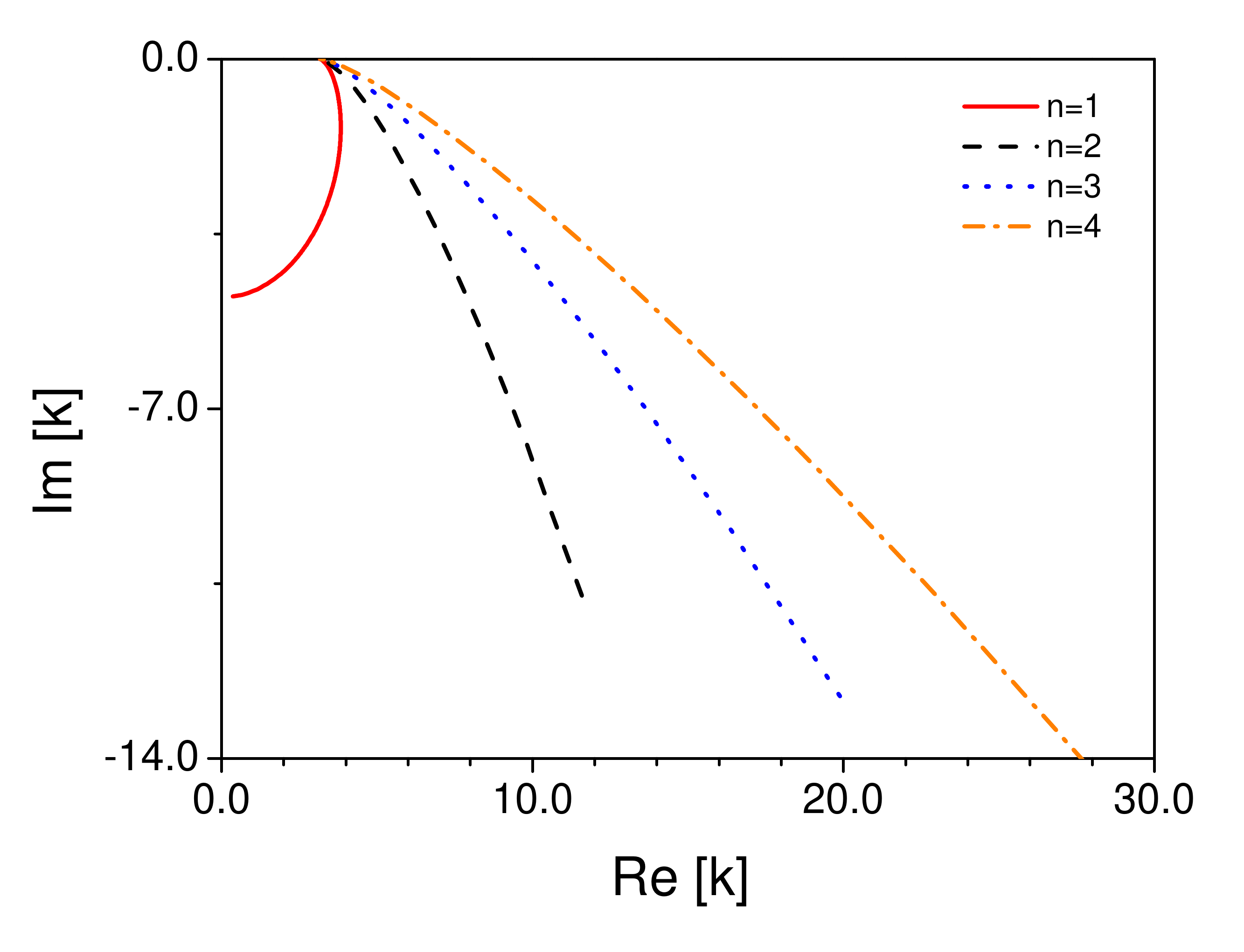}
\caption{Plot of the first four poles $k_n=\alpha_n-i\beta_n$ on the fourth quadrant of the complex $k$ plane for a fixed barrier height $V_0=10$ as a function of the barrier width $L$ in the range from $L=100$, where they are located just above the barrier height $k_0 \approx 3.16$, down to $L=0.42$, which shows that they migrate to higher values of $\alpha_n$ and $\beta_n$, except the first pole. See text.}
\label{Fig3}
\end{figure}
\begin{figure}[htbp!]
\includegraphics[width=3.6in]{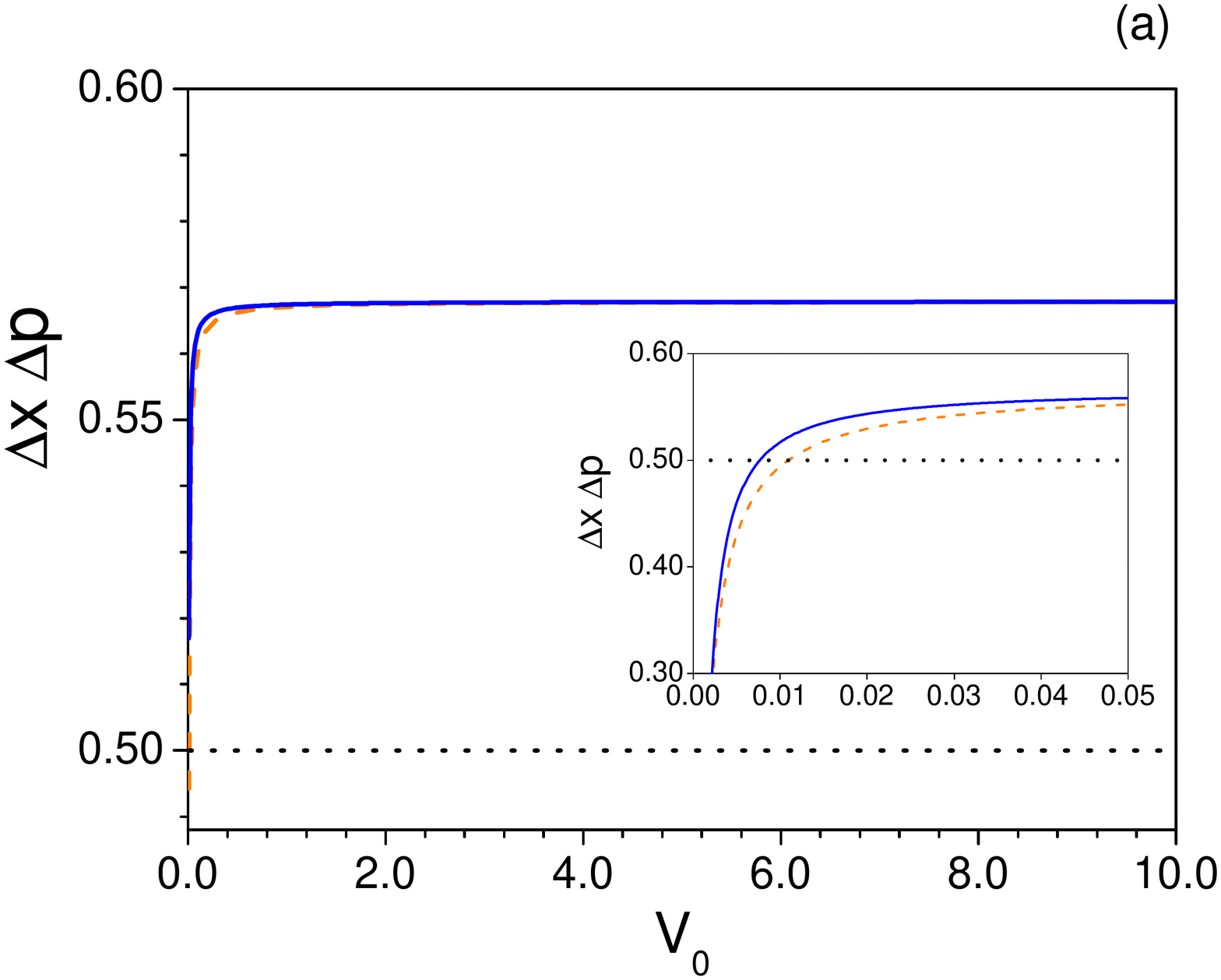}
\includegraphics[width=3.6in]{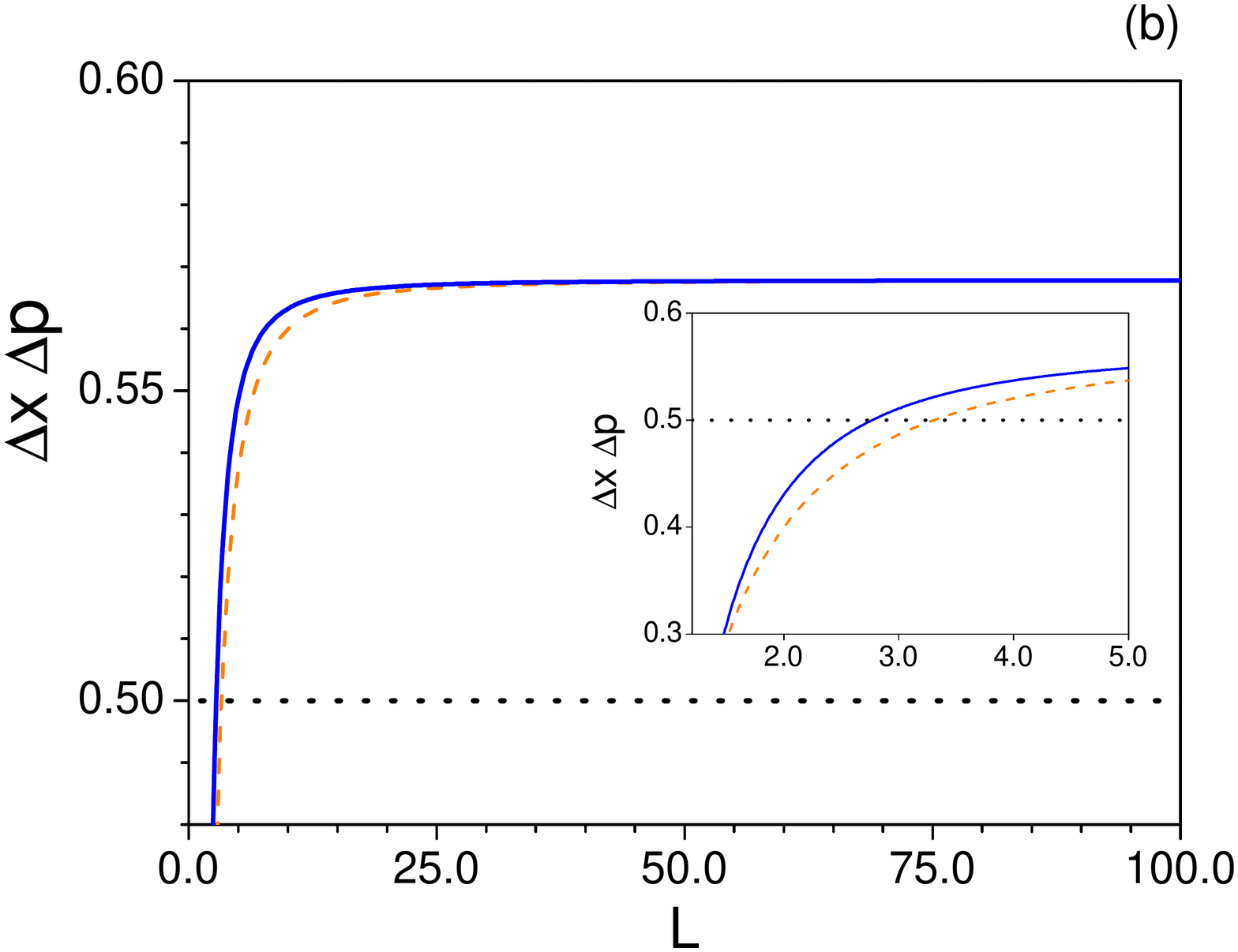}
\caption{ Plot of the uncertainty $\Delta x \,\Delta p$ using Eqs.~(\ref{i1}) (blue solid line) and (\ref{i1B}) (orange dashed line) for a one-dimensional potential barrier: (a) for a fixed value of the barrier width $L=100$ as a function of the barrier height $V_0$  and (b) for a fixed value of the barrier height $V_0=10$ as a function of the barrier width $L$. Both cases refer to the resonance state with $n=1$. See text.}
\label{Fig4}
\end{figure}

The \textit{resonance  state} solutions to Eq. (\ref{f1}) with the potential given by (\ref{1rp}) read,
\begin{equation}
u_n(x)=\left\{
\begin{array}{cc}
F_n \,e^{-ik_nx}, &\,\, \leq 0,\\[.3cm]
A_n \,e^{iq_nx} +B_n \,e^{-iq_nx}&, \,\,\, 0 \leq x \leq L \\[.3cm]
D_n \,e^{ik_nx}, & x \geq L,
\end{array}
\right.
\label{2rp}
\end{equation}
where $q_n=[k_n^2-V_0]^{1/2}$. From the usual continuity conditions of the above solutions at $x=0$ and $x=L$, one obtains the even and odd solutions to the problem. The corresponding complex $k_n$'s satisfy the equations,
\begin{equation}
J_{(\pm)}(k_n)=e^{-iq_n/2}(k_n+q_n) \mp e^{iq_n /2}(k_n-q_n)=0,
\label{2rb}
\end{equation}
with the  $(+)$ sign corresponding to the even and the sign $(-)$ to the odd solutions. The set of complex $k_n$'s may be obtained by methods similar to those employed in Ref.  \cite{nussenzveig59} for the $s$-wave rectangular barrier in three dimensions, as discussed in Ref. \cite{vgc04}.

Figure \ref{Fig3} exhibits a plot in the $k$ plane of the motion of the first four complex poles of the problem for a fixed value of the potential barrier height $V_0=10$ as a function of the barrier width $L$. The range of values of $L$ goes from $L=100$ to $L=0.42$. Around the value  $L=100$, the poles are clustering together just above the barrier height, which in $k$ space corresponds to the wave number $k_0=\sqrt{V_0}\approx 3.16$. As $L$ decreases, the poles move along the $k$ plane increasing its values of $\alpha_n$ and $\beta_n$, except the first complex pole which migrates towards the imaginary $k$ axis, $n=1$, as discussed in Ref. \cite{vgc04}.

Figure \ref{Fig4}, refers to two graphs involving the rectangular potential. In Fig. 4a we plot for the case $n=1$, the Heisenberg uncertainty relations as function of $V_0$, with a fixed value of the barrier width $L=100$, to make a comparison between our prescription (blue solid line) and the regularization procedure (orange dashed line). One sees that both prescriptions behave in a similar fashion and satisfy the Heisenberg uncertainty relations for a wide range of values of $V_0$. Figure 4b exhibits a plot of the Heisenberg uncertainty relations as a function of the barrier width $L$ for a fixed values of the potential height $V_0=10$. As may be appreciated the uncertainty relations are also very similar for both prescriptions and are satisfied for a large range of values of $L$.

To conclude this Section, is worth mentioning that the validity of the Heisenberg uncertainty relations for a broad range of potential parameters is fulfilled as $n$ increases in both potential models.

\section{Conclusions} \label{conclusions}

In this work we have explored the validity of the Heisenberg uncertainty relations for the resonance
solutions to the Schr\"odinger equation for a single particle potential in an exact analytical fashion.
\textit{Resonance  states} constitute a non-Hermitian basis for potentials of arbitrary shape that vanish exactly beyond a distance.  Following the expression of the normalization condition for \textit{resonance  states}, which involves an integral contribution along the internal interaction region plus a surface term, we have considered a definition for the expectation value of an operator  which is also given by the sum of an integral term plus a surface contribution. We have found that the Heisenberg uncertainty relations involving resonance states require that the corresponding complex poles are \textit{proper}, i.e., $\alpha_n > \beta_n$. Otherwise, the resonance energy becomes negative and that implies a negative expectation value for the momentum square which invalidates the uncertainty relations.
We have also made a comparison of our method with the regularization method, and found that both methods yield the same analytical results for the expectation values of $H$, $p$ and $p^2$ but differ in the regularization method, in one and two additional surface terms for the expectation values of $x$ and $x^2$. These terms become only  relevant for poles located very close to the threshold energy. Model calculations show that both procedures give almost the same results and satisfy the Heisenberg uncertainty relations for a broad range of potential parameters.

Notice that for the Hamiltonian to the system, the dispersion given by (\ref{ia1}) yields, in view of (\ref{e3n}),  $\Delta \mathcal{H}=0$, which is consistent with the fact that the particle is described by the eigenfunction $u_n(r)$. An issue of interest for future work is to consider the expectation value of an operator involving an arbitrary wave function $\Psi(r)$ that may be expanded in terms of resonance states to address the issue of measurement from a non-Hermitian perspective.

The present work shows that the Heisenberg uncertainty relations may hold beyond the standard Hermitian framework of quantum mechanics. This might be of particular interest for those pursuing a line of inquiry that explores the possibility of extending the standard formalism of quantum mechanics to incorporate in a fundamental fashion a non-Hermitian treatment of the Hamiltonian to the system, as suggested by studies on tunneling decay \cite{gcmv12,gcmv13}.

\begin{acknowledgments}
G.G.-C. and J.V. acknowledge partial financial support of DGAPA-UNAM-PAPIIT IN105216 and IN105618, Mexico, and J.V. from UABC under Grant PROFOCIE-2017.
\end{acknowledgments}
\appendix
\section{Regularization procedure for $\braket{r^m}_B$}\label{appendixA}

We derive here a useful mathematical identity starting from the following identity presented in the works of Berggren\cite{berggren68} and  Gyarmati\cite{gyarmati71}:
\begin{equation}
\lim_{\varepsilon\to 0} \int_0^{\infty}e^{-\varepsilon r^2}\,e^{z r}\,dr=-\frac{1}{z},
\label{identity0}
\end{equation}
where according to Ref. \cite{berggren68}, the result holds for $Re(z)>0$, and $Re(z)<0$.
By computing the $m$-th partial derivative of  the result of Eq. (\ref{identity0}) with respect to the variable $z$,  we obtain
\begin{eqnarray}
\lim_{\varepsilon\to 0} \int_0^{\infty}e^{-\varepsilon r^2}\,r^m\,e^{z r}\,dr=\frac{\partial^m}{\partial z^m}\left(-\frac{1}{z}\right).
\label{identity1}
\end{eqnarray}
The above result then can be written as,
\begin{eqnarray}
\int_0^{a} \,r^m\,e^{z r}\,dr+ \lim_{\varepsilon\to 0} \int_a^{\infty}e^{-\varepsilon r^2}\,r^m\,e^{z r}\,dr =  \nonumber \\ [.3cm]
\frac{\partial^m}{\partial z^m}\left(-\frac{1}{z}\right).
\label{talacha1}
\end{eqnarray}

The first integral in (\ref{talacha1})  may also be written as the  $m$-th partial derivative (with respect to the parameter $z$),
\begin{eqnarray}
&&\lim_{\varepsilon\to 0} \int_a^{\infty}e^{-\varepsilon r^2}\,r^m\,e^{z r}\,dr =  \nonumber \\ [.3cm]
&&\frac{\partial^m}{\partial z^m}\left(-\frac{1}{z}\right)-\frac{\partial^m}{\partial z^m} \left(\int_0^a\, e^{zr}\,dr \right)=  \nonumber \\ [.3cm]
&&{\frac{\partial^m}{\partial z^m}\left(-\frac{1}{z}\right)}-\frac{\partial^m}{\partial z^m} \left(\frac{e^{za}}{z}{-\frac{1}{z}} \right),
\label{talacha1a}
\end{eqnarray}
so we may finally write the above identity as,
\begin{equation}
\lim_{\varepsilon\to 0} \int_a^{\infty}e^{-\varepsilon r^2}\,r^m\,e^{z r}\,dr=-\frac{\partial^m}{\partial z^m} \left(\frac{e^{za}}{z} \right).
\label{identity2}
\end{equation}

\section{Regularization procedure for $\braket{p^m}_B$} \label{appendixB}

Let us define in Eq.~(\ref{b3}) $\mathcal{O} \equiv p^m$, with $p=-i\,d/dr$,  and use Zel'dovich's regularization procedure
\begin{eqnarray}
\braket{p^m}_B&=&\frac{1}{i^m}\int_0^{a} u_n \frac{d^mu_n(r)}{dr^m} \,\,dr \nonumber \\ [.3cm]
&+&\lim_{\varepsilon \to 0}\frac{1}{i^m}\int_a^{\infty} e^{-\varepsilon r^2}\,u_n \frac{d^m u_n(r)}{dr^m}\,dr.
\label{apsB_1}
\end{eqnarray}
Since along the external region ($r>a$), $u_n(r)=D_n{\rm e}^{i k_n r}$, we compute the $m$-th derivative $d^mu_n(r)/dr^m$, and write,
\begin{eqnarray}
\braket{p^m}_B&=&\frac{1}{i^m}\int_0^{a} u_n \frac{d^mu_n(r)}{dr^m} \,\,dr \nonumber \\ [.3cm]
&+&D_n^2 k_n^m\,\lim_{\varepsilon \to 0}\,\int_a^{\infty} e^{-\varepsilon r^2}\,e^{z r} \,dr,
\label{apsB_2}
\end{eqnarray}
with $z=2 i k_n$. The second integral in Eq.~(\ref{apsB_2}) can be readily evaluated by using the identity derived from Eq.~(\ref{identity1}), namely
\begin{equation}
\lim_{\varepsilon\to 0} \int_a^{\infty}e^{-\varepsilon r^2}\,e^{z r}\,dr=-\frac{e^{za}}{z},
\label{identity_B2}
\end{equation}
and obtain,
\begin{equation}
\braket{p^m}_B=\frac{1}{i^m}\int_0^{a} u_n \frac{d^mu_n(r)}{dr^m} \,\,dr
+\frac{i}{2} k_n^{m-1}\,u_n^2(a).
\label{apsB_3}
\end{equation}

In what follows, let us calculate from  Eq.~(\ref{apsB_3}) the expectation values $\braket{p}_B$, and $\braket{p^2}_B$.
For the case  $\braket{p}_B$,  let us choose $m=1$ in  Eq.~(\ref{apsB_3}), which leads to,
\begin{eqnarray}
\braket{p}_B&=&\frac{1}{i}\int_0^{a} u_n \frac{du_n(r)}{dr} \,\,dr +\frac{i}{2} \,u_n^2(a), \nonumber \\ [.3cm]
&=&\frac{1}{i}\int_0^{a} \frac{1}{2} \frac{du_n^2(r)}{dr}\,\,dr +\frac{i}{2} \,u_n^2(a), \nonumber \\ [.3cm]
&=&\frac{1}{i} \frac{1}{2} \left [ u_n^2(r) \right ]_{0}^{a} +\frac{i}{2} \,u_n^2(a) =0,
\label{apsB_4}
\end{eqnarray}
and hence, using Eq. (\ref{b4}), it follows that $\braket{\braket{p}}_B=0$.

For the case  $\braket{p^2}_B$ let us choose $m=2$ in  Eq.~(\ref{apsB_3}),
\begin{equation}
\braket{p^2}_B=-\int_0^{a} u_n(r) \frac{d^2 u_n(r)}{dr^2} \,\,dr
+\frac{i}{2} k_n\,u_n^2(a).
\label{apsB_5}
\end{equation}
By using  the relation from Eq.~(\ref{f1a}), $-d^2/dr^2=H-V(r)$, and $H u_n(r)= E_n u_n(r)$ where we recall that $E_n={\cal E}_n-i\Gamma_n/2$, we obtain
\begin{equation}
\braket{p^2}_B=E_n \int_0^{a} u_n^2(r) \,\,dr - \int_0^{a} V(r)\,u_n^2(r) \,\,dr +\frac{i}{2} k_n\,u_n^2(a).
\label{apsB_6}
\end{equation}
We evaluate the first integral in the right hand side of Eq.~(\ref{apsB_6}) by using the normalization condition (Eq.~(\ref{f4})) in the form,
\begin{equation}
\int_0^au_n^2(r)dr =1- \frac{i}{2k_n}u_n^2(a),
\label{f4_B}
\end{equation}
which allows us to write  Eq.~(\ref{apsB_6})  as,
\begin{equation}
\braket{p^2}_B=E_n- \int_0^{a} V(r)\,u_n^2(r) \,\,dr,
\label{apsB_8}
\end{equation}
and hence, using again Eq. (\ref{b4}),  $\braket{\braket{p^2}}_B= {\rm Re}\braket{p^2}_B$.

\end{document}